# MPPC versus MRS APD in two-phase Cryogenic Avalanche Detectors

A. Bondar,[a,b] A. Buzulutskov,[a,b] A. Dolgov,[b] E. Shemyakina,[a,b,*] A. Sokolov,[a,b]

[a] *Budker Institute of Nuclear Physics SB RAS, Lavrentiev avenue 11, 630090 Novosibirsk, Russia*
[b] *Novosibirsk State University, Pirogov street 2, 630090 Novosibirsk, Russia*

*E-mail:* A.F.Buzulutskov@inp.nsk.su

ABSTRACT: Two-phase Cryogenic Avalanche Detectors (CRADs) with combined THGEM/GAPD multiplier have become an emerging potential technique for dark matter search and coherent neutrino-nucleus scattering experiments. In such a multiplier the THGEM hole avalanches are optically recorded in the Near Infrared (NIR) using a matrix of Geiger-mode APDs (GAPDs). To select the proper sensor, the performances of six GAPD types manufactured by different companies, namely by Hamamatsu (MPPCs), CPTA (MRS APDs) and SensL (SiPMs), have been comparatively studied at cryogenic temperatures when operated in two-phase CRADs in Ar at 87 K. While the GAPDs with ceramic packages failed to operate properly at cryogenic temperatures, those with plastic packages, namely MPPC S10931-100P and MRS APD 149-35, showed satisfactory performances at 87 K. In addition, MPPC S10931-100P turned out to be superior in terms of the higher detection efficiency, lower nose rate, lower pixel quenching resistor and better characteristics reproducibility.

---

[*] Corresponding author.

**Contents**

**1. Introduction**

Two-phase Cryogenic Avalanche Detectors (CRADs) with THGEM and THGEM/GAPD multipliers have become an emerging potential technique for dark matter search and coherent neutrino-nucleus scattering experiments [1] and for their energy calibration [2]. In two-phase CRADs the primary ionization charge, composed of electrons produced in the noble-gas liquid and emitted into the gas phase, can be multiplied in the gas phase with a THGEM charge multiplier [3],[4],[5],[6]. In addition, in a combined THGEM/GAPD charge/optical multiplier, the THGEM hole avalanches are optically recorded either in the Near Infrared (NIR) [7],[8],[9] or the Vacuum Ultraviolet (VUV) [10],[11] using a matrix of Geiger-mode APDs (GAPDs, [12]), operated at cryogenic temperatures. Such a combined charge/optical readout would result in a higher overall gain at superior spatial resolution [1],[8].

To provide such a readout in the two-phase CRAD in Ar with an active volume of the order of 100 l, it was proposed to employ as many as a thousand of GAPD sensors operated at 87 K [13]. At the moment such a project is being developed in the Budker INP and Novosibirsk State University [13],[14],[15]. Accordingly, it is important to make a proper choice of the GAPD sensor, which should have a higher efficiency in the NIR, higher gain, lower noise rate and higher performance reliability at cryogenic temperatures. In addition it is desirable to have a lower GAPD pixel quenching resistor, to prevent the effect of the GAPD performance degradation at cryogenic temperatures observed elsewhere [16]. It should be remarked that though the GAPD operation at cryogenic temperatures was repeatedly studied [17],[18],[19],[20],[21],[22],[23],[24],[25], the understanding of their performance at low temperatures is still incomplete.

In the present work the proper GAPD sensor was selected between those produced by three companies, namely between MPPCs (Multi-Pixel Photon Counters, produced by Hamamatsu [26]), MRS APDs (Metal Resistor Semiconductor APDs, produced by CPTA [27]) and SiPMs (Silicon Photo-Multipliers, produced by SensL [28]). Their performances have been comparatively studied when operated in two-phase CRADs in Ar at 87 K. Gain, nose rate and pixel resistance characteristics and the relative detection efficiency in the NIR have been

compared for the following GAPD types: MPPC S10931-100P, MPPC S10362-33-100C, MRS APD 149-35, MRS APD 150-50, MRS APD 140-40 and SiPM MicroSM-30035-X13.

## 2. Experimental setup

The experimental setup was similar to those used in our previous measurements with two-phase CRADs in Ar [8],[16]. It includes a 9 l cryogenic chamber, comprising a double-THGEM multiplier with an active area of $10\times10$ cm$^2$ placed within the saturated vapour above the liquid: see Fig. 1. The cathode electrode was immersed in liquid Ar with an active liquid layer thickness of 5 mm. The detector was operated in two-phase Ar in equilibrium state, at a saturated vapour pressure of 1.0 atm corresponding to a temperature of 87 K.

There were two measurement sessions with different GAPD assemblies placed in the gas phase behind the second THGEM. In the first session the GAPD assemblies included three GAPD types of CPTA production: these were MRS APDs 149-35, 150-50 and 140-40, with an active area of $2.1\times2.1$, $2.5\times2.5$ and $3\times3$ mm$^2$ respectively, in plastic packages [27].

In the second session the GAPD assembly consisted of seven GAPD samples (see Fig. 2); it was placed at a distance of 14 mm from the second THGEM, electrically insulated from the latter by an acrylic plate transparent in the NIR and by a wire grid at ground potential (see Fig. 1). Four GAPD types were studied in the second session; their characteristic properties are presented in Table 1. These were MPPCs S10931-100P (sample #1) and S10362-33-100C (samples #2 and #3) with an active area of $3\times3$ mm$^2$ in plastic and ceramic packages respectively, MRS APDs 149-35 (samples #5 and #6) with an active area of $2.1\times2.1$ mm$^2$ in plastic packages and SiPM MicroSM-30035-X13 (sample #7) with an active area of $3\times3$ mm$^2$ in ceramic package.

The GAPD signals were read out via 1 m long twisted pair cables connected to fast amplifiers (produced by CPTA [27]) with a 300 MHz bandwidth and amplification factor of 30. A TDS5032B digital oscilloscope serves as a data acquisition system.

The details of the experimental procedures regarding determination of the GAPD noise rate, gain and pixel resistance, can be found elsewhere [7],[16],[23]. In particular, the noise rate was measured by counting the noise pulses in a given time interval, the noise signals being recognized by their characteristic pulse-shapes. The GAPD gain was measured using the noise signals: the appropriate procedure included the determination of the pulse-area for the single-pixel signal, at a given fast amplifier gain, which in turn permitted to calculate the single-pixel charge (in fact equal to the GAPD gain) at a given bias voltage.

To measure the relative detection efficiency of GAPD samples in the NIR, the detector was irradiated from outside through a steel collimator and aluminium windows by 15-40 keV X-rays from a pulsed X-ray tube (at a pulse rate of 240 s$^{-1}$): see Fig. 1. The GAPDs recorded NIR photons from avalanche scintillations in the THGEM holes, induced by X-ray absorption in the liquid layer, followed by emission of ionization electrons into the gas phase and their further multiplication in the THGEM holes. Due to a relatively large distance between the GAPD assembly and the second THGEM (14 mm), the GAPD samples of the assembly were irradiated practically uniformly by the NIR photon flux. The detection efficiency of the particular GAPD was measured by counting the single- or multiple-pixel pulses in the GAPD signal, which is equivalent to counting the number of single-photoelectron (p.e.) pulses in the signal per X-ray pulse. This allowed one to compare the detection efficiency of different GAPDs at a given NIR photon flux. Note that the detection efficiency measured that way (the "overall" efficiency) is proportional to the GAPD active area.

More details on the detection principles of gaseous Ar scintillations in the NIR can be found elsewhere [1],[13],[29]. In particular Fig. 3 presents the absolute Photon Detection Efficiency (PDE) of GAPDs of the MPPC S10931-100P, MPPC S10362-33-100C and CPTA 149-35 types, taken or deduced from those of the producer web sites [26],[27], along with the scintillation emission spectrum of gaseous Ar in the NIR [30].

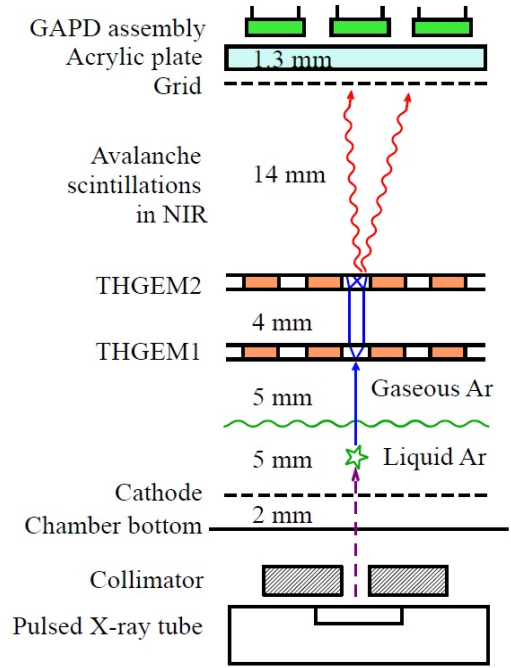

Fig. 1. Experimental setup to study the GAPD performances in two-phase CRADs in Ar.

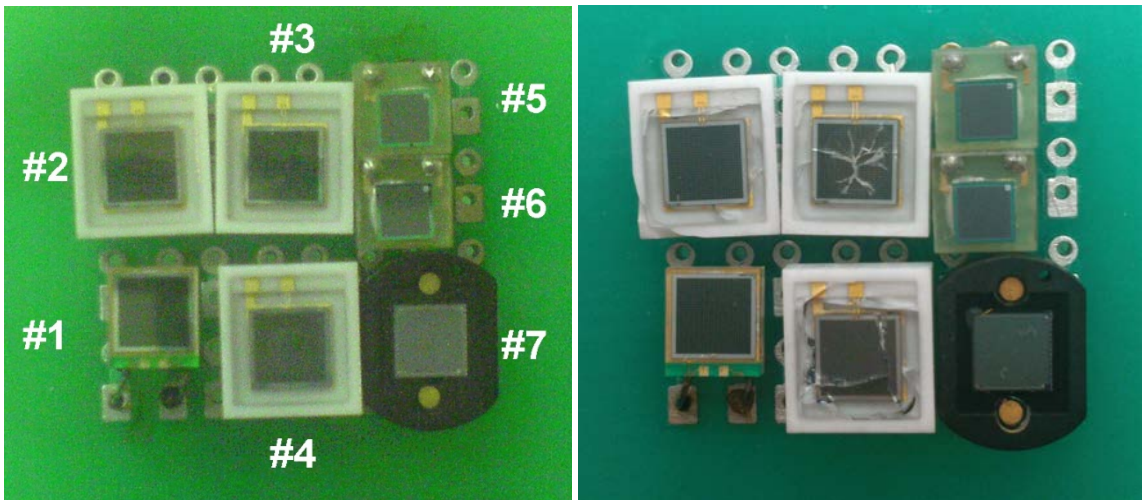

Fig. 2. Photograph of the GAPD assembly in the second measurement session before (left) and after (right) the cryogenic measurement. The GAPD sample numbers used throughout the paper are indicated in the left panel.

| GAPD sample # | GAPD type (producer) | Package material | Active area, mm² | Number of pixels | Active area fill factor, % | Typical bias voltage at 295 K, V | Typical bias voltage at 87 K, V |
|---|---|---|---|---|---|---|---|
| 1 | MPPC S10931-100P (Hamamatsu) | plastic | 3×3 | 900 | 78.5 | 71 | 62 |
| 2 | MPPC S10362-33-100C (Hamamatsu) | ceramic | 3×3 | 900 | 78.5 | 70 | 61 |
| 3 | | | | | | 70 | 61 |
| 4 | | | | | | 72 | failed |
| 5 | MRS APD 149-35 (CPTA) | plastic | 2.1×2.1 | 1764 | 62 | 40 | 36 |
| 6 | | | | | | 40 | 36 |
| 7 | SiPM MicroSM-30035-X13 (SensL) | ceramic | 3×3 | 4774 | 64 | 27 | failed |

Table 1. Characteristic properties of GAPDs studied in the second measurement session [26],[27],[28]. The last two columns present the typical bias voltages at which the GAPD single-pixel pulse-height was of the order of 5 mV, after the fast amplifier at 50 Ohm load resistance.

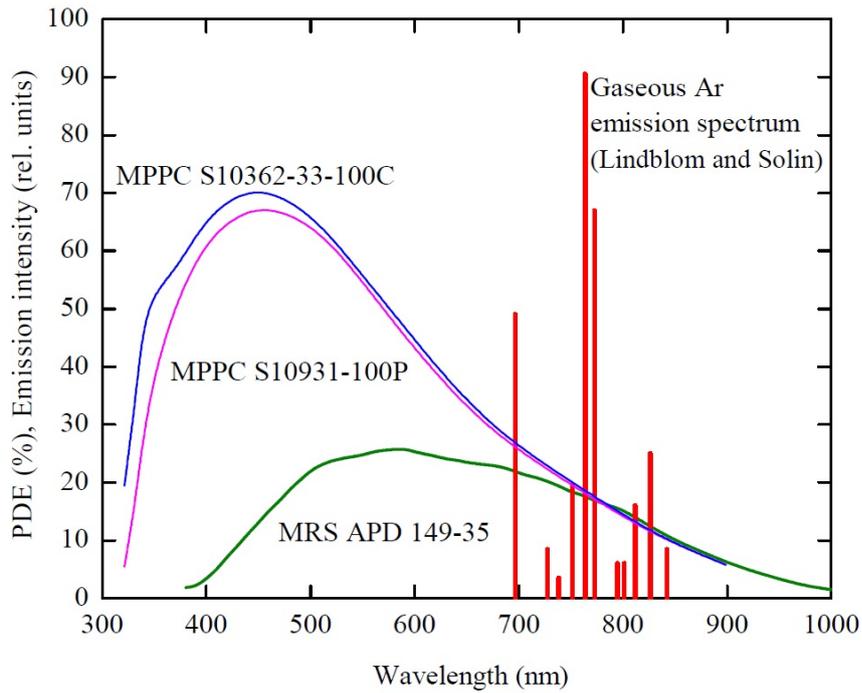

Fig. 3. Gaseous Ar scintillation emission spectrum in the NIR [30] and those of the absolute Photon Detection Efficiency (PDE) of GAPDs of the MPPC S10931-100P [26], MPPC S10362-33-100C [26] and CPTA 149-35 [1],[27] types.

## 3. Results

In the first measurement session, the proper sensor was selected from those produced by CPTA, using gain and noise rate characteristics measured at 87 K. Fig. 4 presents the appropriate gain-voltage characteristics, namely the GAPD gain (the single pixel charge expressed in electrons) as a function of the bias voltage, while Fig. 5 presents the noise rate as a function of the bias voltage. Comparing Figs. 4 and 5, one may conclude that MRS APD 149-35, having a reasonable active area, of $2.1 \times 2.1$ mm$^2$, have a superior performance in terms of the maximum gain and minimum noise rate. In particular at an overvoltage of 14 V its gain reached a value of $1.5 \times 10^6$, which is a factor of 3 higher than the maximum gains obtained for MRS APDs of larger active area. In addition, its noise rate at this maximum gain, $<10^3$ s$^{-1}$, is an order and two orders of magnitude lower compared to MRS APDs with an active area of $2.5 \times 2.5$ and $3 \times 3$ mm$^2$ respectively (at their maximum gains). Accordingly in the second measurement session the CPTA-made GAPDs were represented by those of the MRS APD 149-35 type.

In the second measurement session, the GAPD performances were first studied at room temperature and then at cryogenic temperature, at 87 K in the two-phase mode. As seen from Fig. 2 and Table 1, the GAPDs with ceramic packages failed to operate properly at cryogenic temperatures: their glass windows cracked after the first cryogenic measurement, apparently due to the different thermal expansions of the glass window and the ceramic package. One half of those devices completely lost their signals while another half continued to operate even with the windows cracked. Such a situation with the GAPDs with ceramic packages cannot be considered normal.

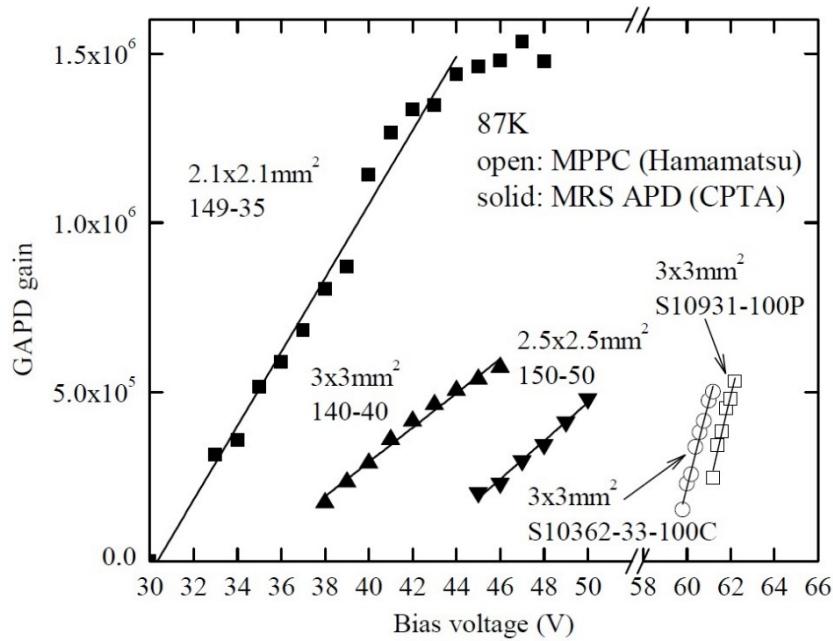

Fig. 4. Typical gain-voltage characteristics of different GAPD types at 87 K, namely of three types of MRS APDs (produced by CPTA) and two types of MPPCs (produced by Hamamatsu). The appropriate GAPD type names and active areas are indicated in the figure.

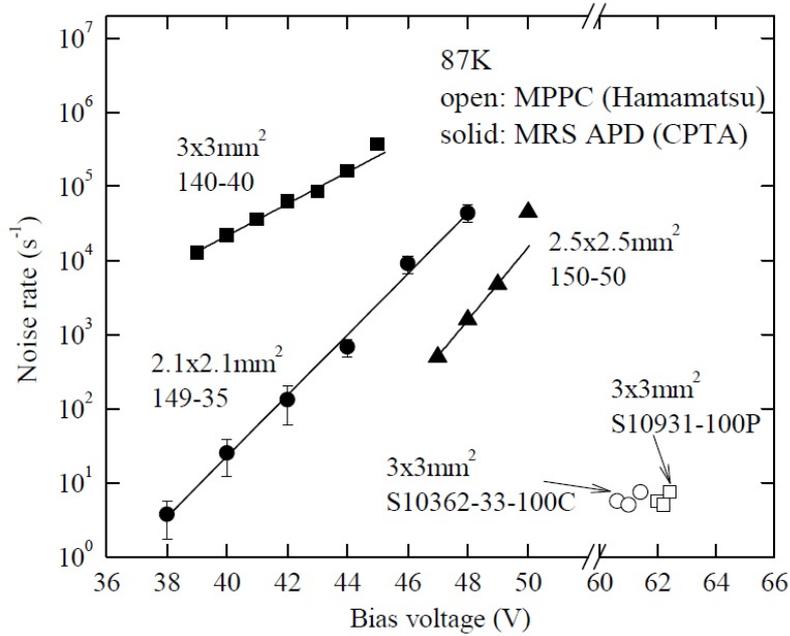

Fig. 5. Typical noise rates of different GAPD types as a function of the bias voltage, at 87 K. The appropriate GAPD type names and active areas are indicated in the figure.

In contrast, the GAPDs with plastic packages, namely those of the MPPC S10931-100P and MRS APD 149-35 type, showed satisfactory performance at cryogenic temperatures: they have no damage problems even after multiple cryogenic runs.

As seen from Table 1, the GAPD operating voltages significantly decrease with decreasing temperature, in particular for MPPCs by 10 V. It is interesting that MPPCs have a rather narrow range of operating voltages as compared to MRS APDs, of only 1 V for MPPC S10931-100P versus 15 V for MRS APD 149-35 (see Fig. 4), the maximum voltage reaching a value of 62 V for the former. At voltages higher than the maximum, the MPPC performance became unstable in terms of enhanced cross-talks, after-pulses and noise rates.

Though the maximum gain of MPPCs, of $5\times10^5$, is a factor 3 lower compared to that of MRS APDs (see Fig. 4), it is high enough to effectively record and digitize the signals even in a single-photoelectron (i.e. single-pixel) counting mode. This is clearly seen in Fig. 6 showing typical single-pixel pulses of MPPC S10931-100P (accompanied by multi-pixel pulses and an after-pulse) and those of MRS APD 149-35, at 87 K. One can see that at the maximum operating voltage (at a gain of $5\times10^5$), the typical single-pixel pulse-height of MPPC (2 mV) is somewhat smaller while the signal width (120 ns) is substantially larger, as compared to those of MRS APD at the same gain (9 mV and 15 ns respectively). One can also see from Fig. 5 that MPPCs are superior in terms of the minimal noise rate: at the maximum operating voltages their noise rate was as low as few Hz. Note that the contribution of the after-pulses at the maximum voltage was inessential, below 15%.

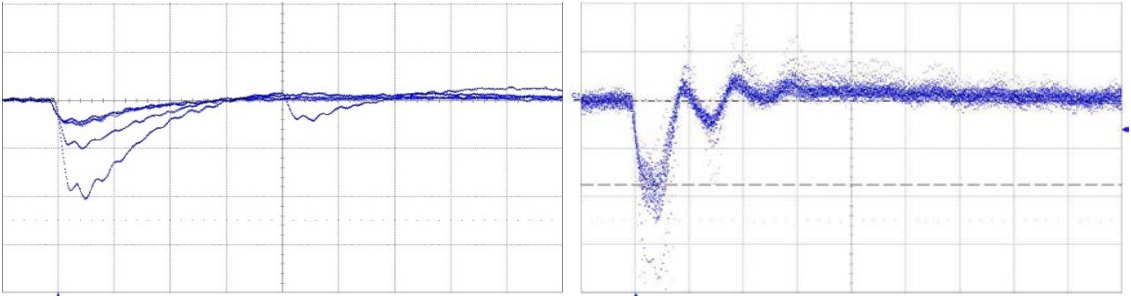

Fig. 6. Typical GAPD single-pixel pulses in two-phase CRADs in Ar at 87 K, after the fast amplifier at 50 Ohm load resistance: for MPPC S10931-100P at a voltage of 62 V (left, 50 ns/div; 5 mV/div) and MRS APD 149-35 at a voltage of 38 V (right, 20 ns/div; 5 mV/div).

Recently the effect of the GAPD performance degradation at cryogenic temperatures, at higher incident photon fluxes, has been observed for GAPDs of the MRS APD type [16]. The critical counting rate of photoelectrons produced at MRS APD 149-35 to degrade its performance at 87 K, was estimated to be of the order of $10^4$ s$^{-1}$. It was shown to result from the increase of the pixel recovery time $\tau = R_Q C_P$ due to dramatically increase of the pixel quenching resistor ($R_Q$) at 87 K, sometimes reaching a value of 40 G$\Omega$. Accordingly it is desirable to have the GAPD pixel resistor as lower as possible.

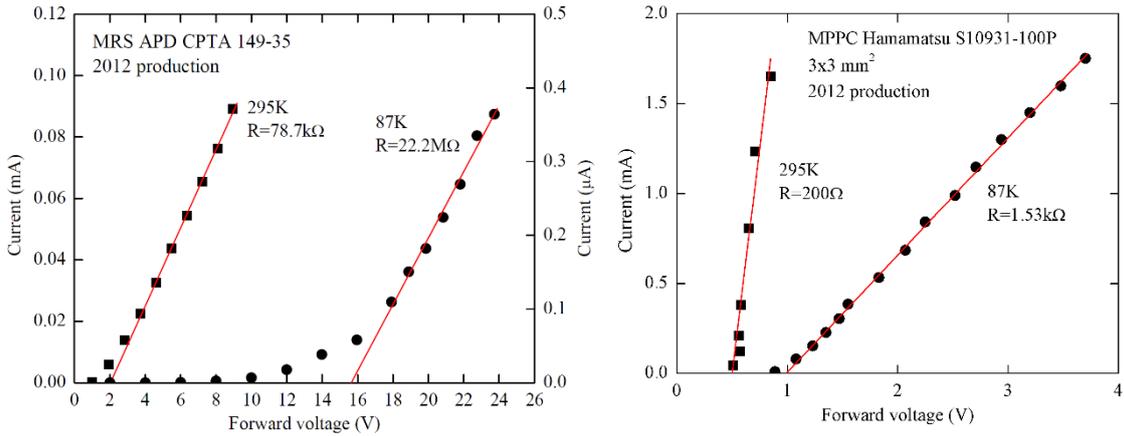

Fig. 7. GAPD current-voltage characteristics in the forward direction at room temperature (squares) and at 87 K (circles). Left: MRS APD 149-35 (left scale corresponds to 295 K, right scale to 87 K). Right: MPPC S10931-100P. The slope of the linear part of the I-V curve is defined by the GAPD total quenching resistor R.

Fortunately MPPCs meet this requirement, in contrast to MRS APDs. This is seen from Fig. 7 presenting the current-voltage characteristics in the forward direction, at room and cryogenic temperature. From those one can calculate the GAPD total quenching resistor (R) using the slope of the linear part of the I-V curve. The pixel quenching resistor is equal to $R_Q = R \times N_P$, where $N_P$ is the number of pixels for a given GAPD type; the appropriate $R_Q$ values at room temperature and 87 K are presented in Table 2. One can see that the MPPC pixel quenching resistor at 87 K is more than four orders of magnitude lower than that of MRS APD, which apparently solve the problem of the performance degradation at higher photon fluxes.

It should be remarked that the quenching-resistor values of MRS APDs varied considerably within and between the production batches [16]: by a factor of 2 within the batch and by a factor of 4-7 between different batches, in particular between those of 2009 and 2012 production. Such variations in the production characteristics of CPTA-made MRS APDs are rather disturbing. In contrast, the MPPC characteristics were fairly well reproducible within and between the production batches, in our experience.

| GAPD type | $R_Q$ at 295 K | $R_Q$ at 87 K |
|---|---|---|
| MPPC S10931-100P | 180 kΩ | 1.4 MΩ |
| MRS APD 149-35 | 140 MΩ | 39 GΩ |

Table 2. Pixel quenching resistor ($R_Q$) at room temperature and at 87 K for two GAPD samples of Fig. 7.

The relative detection efficiency of different GAPD samples at a given NIR photon flux were measured by counting the number of single- and multiple-pixel pulses in the GAPD signal per X-ray pulse, as described in section 2. Examples of such signals are shown in Fig. 8 for MPPC S10931-100P and MRS APD 149-35. The rather large signal width, of tens of microseconds, is due to the slow electron emission component presented in two-phase Ar systems [1]. The NIR photon flux was defined by the X-ray pulse intensity and the THGEM gain; it was kept below the critical value, to prevent the performance degradation of MRS APD 149-35.

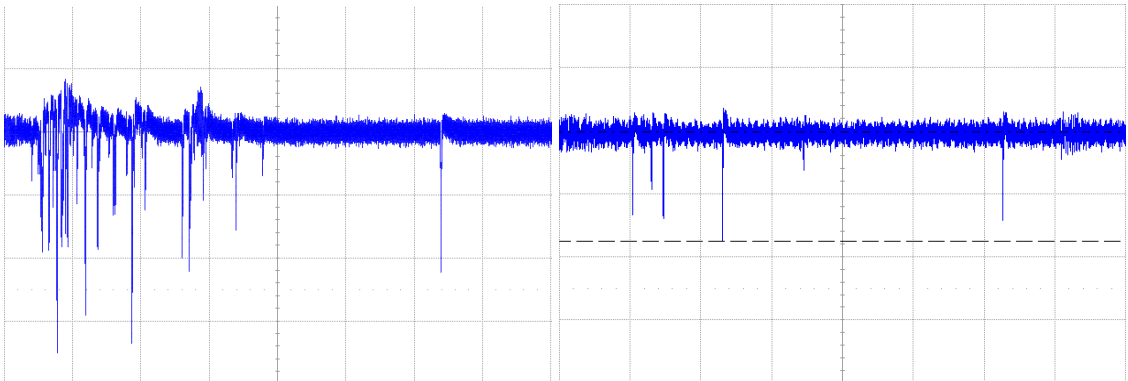

Fig. 8. Typical GAPD signals per X-ray pulse in the two-phase CRAD in Ar at 87 K, induced by avalanche scintillations from the THGEM holes in the NIR: for MPPC S10931-100P (left) and MRS APD 149-35 (right). Scales: 2 µs/div; 5 mV/div.

The efficiency values are presented in Table 3 for different GAPD samples of three GAPD types. For all the GAPD samples of the MPPC type, no matter with ceramic or plastic packages, the relative efficiencies are close to each other within the measurement uncertainties, at the maximum bias voltages. This is in accordance with the absolute PDE spectra of these MPPC types presented in Fig. 3: the spectra are practically identical, in particular in the NIR. On the other hand, the relative efficiencies of the samples of the MRS APD type turned out to be substantially reduced compared to those of the MPPC type, by a factor of 4-6. The factor of 2 of this reduction is explained by a smaller active area: 4.4 mm$^2$ for MRS APD 149-35 versus 9 mm$^2$ for MPPC S10931-100P. The remaining factor, of 2-3, can be explained by the lower absolute PDE of MRS

APD as compared to MPPC, although this somewhat contradicts the data presented by the producers (see Fig. 3). It should be remarked here that we have more confidence in the absolute PDE data of Hamamatsu than CPTA. Anyway, the results obtained clearly speaks in favor of MPPCs when making a choice of the sensor with the highest overall detection efficiency in the NIR, in the emission region of gaseous Ar scintillations.

| Sample # | GAPD type, active area | Bias voltage, V | Average number of p.e. peaks per X-ray pulse | Relative detection efficiency in the NIR |
|---|---|---|---|---|
| 1 | MPPC S10931-100P, 3×3 mm$^2$ | 62 | 37±3 | 1.0±0.1 |
| 2 | MPPC S10362-33-100C, 3×3 mm$^2$ | 60 | 17±2 | 0.46±0.05 |
| 2 | | 61 | 30±3 | 0.81±0.1 |
| 3 | | 61 | 30±3 | 0.81±0.1 |
| 5 | MRS APD 149-35, 2.1×2.1 mm$^2$ | 38 | 6±1 | 0.16±0.02 |
| 6 | | 38 | 7±1 | 0.20±0.02 |

Table 3. Relative detection efficiency in the NIR for different GAPD samples at 87 K, deduced from the average number of photoelectron (p.e.) peaks in the GAPD signal per X-ray pulse.

## 4. Conclusions

In the present work the proper GAPD sensor was selected between MPPCs (Hamamatsu), MRS APDs (CPTA) and SiPMs (SensL), namely between the following GAPD types: MPPC S10931-100P, MPPC S10362-33-100C, MRS APD 149-35, MRS APD 150-50, MRS APD 140-40 and SiPM MicroSM-30035-X13. Their performances have been comparatively studied when operated in two-phase CRADs in Ar at 87 K. Gain, nose rate and pixel resistance characteristics and the relative detection efficiency in the NIR were compared.

While the GAPDs with ceramic packages failed to operate properly at cryogenic temperatures due to input window cracking, those with plastic packages, namely MPPC S10931-100P and MRS APD 149-35, showed satisfactory performances at 87 K.

Compared with other types of MRS APDs, MRS APD 149-35 had the better performance in terms of the maximum gain and minimum noise rate. On the other hand, its quenching-resistor values varied considerably within and between the production batches which is rather disturbing.

Finally, MPPC S10931-100P turned out to be superior in terms of the lower nose rate, lower pixel quenching resistor, better characteristics reproducibility and higher overall detection efficiency in the NIR (by a factor of 5 higher compared to that of MRS APD 149-35).

The results of the present study have been already used in our laboratory when choosing the proper sensor for two-phase CRADs with THGEM/GAPD-matrix multiplier. In particular, as many as 256 sensors of the MPPC S10931-100P type were purchased by us from Hamamatsu and a 5×5 matrix of this GAPD type is currently under study in the two-phase CRAD in Ar.

## 5. Acknowledgements

We are grateful to A. Chegodaev, V. Nosov and R. Snopkov for technical support. This work consisted of the two independent studies separated in time, namely of the first and second measurement sessions conducted on different experimental setups. The first study was supported in part by the grants of the Government of Russian Federation (11.G34.31.0047) and Russian Foundation for Basic Research (15-02-01821). The second study was supported by Russian Science Foundation (project N 14-50-00080).

## References

[1] A. Buzulutskov, *Advances in Cryogenic Avalanche Detectors*, 2012 JINST 7 C02025.

[2] A. Bondar et al., *Measurement of the ionization yield of nuclear recoils in liquid argon at 80 and 233 keV*, Europhys. Lett. 108 (2014) 12001.

[3] A. Breskin et al., *A concise review on THGEM detectors*, Nucl. Instrum. Meth. A 598 (2009) 107, and references therein.

[4] A. Bondar et al., *Thick GEM versus thin GEM in two-phase argon avalanche detectors*, 2008 JINST 3 P07001 [arXiv:0805.2018].

[5] A. Badertscher et al., *First operation of a double phase LAr Large Electron Multiplier Time Projection Chamber with a two-dimensional projective readout anode*, Nucl. Instrum. Meth. A 641 (2011) 48 [arXiv:1012.0483].

[6] A. Bondar et al., *Two-phase Cryogenic Avalanche Detectors with THGEM and hybrid THGEM/GEM multipliers operated in Ar and Ar+N2*, 2013 JINST 8 P02008.

[7] A. Bondar et al., *Direct observation of avalanche scintillations in a THGEM-based two-phase Ar avalanche detector using Geiger-mode APD*, 2010 JINST 5 P08002 [arXiv:1005.5216].

[8] A. Bondar et al., *First demonstration of THGEM/GAPD-matrix optical readout in two-phase Cryogenic Avalanche Detector in Ar*, Nucl. Instrum. Meth. A 732 (2013) 213 [arXiv:1303.4817].

[9] A. Bondar et al, *On the low-temperature performances of THGEM and THGEM/G-APD multipliers in gaseous and two-phase Xe*, 2011 JINST 6 P07008 [arXiv:1103.6126].

[10] P.K. Lightfoot et al., *Optical readout tracking detector concept using secondary scintillation from liquid argon generated by a thick gas electron multiplier*, 2009 JINST 4 P04002.

[11] D. Yu. Akimov et al., *Two-phase xenon emission detector with electron multiplier and optical readout by multipixel avalanche Geiger photodiodes*, 2013 JINST 8 P05017.

[12] D. Renker and E. Lorenz, *Advances in solid state photon detectors*, 2009 JINST 4 P04004, and references therein.

[13] A. Bondar et al., *Study of infrared scintillations in gaseous and liquid argon. Part II: light yield and possible applications*, 2012 JINST 7 P06014 [arXiv:1204.0580].

[14] A.E. Bondar et al., *PROPOSAL FOR TWO-PHASE CRYOGENIC AVALANCHE DETECTOR FOR DARK MATTER SEARCH AND LOW-ENERGY NEUTRINO DETECTION*, Novosibirsk State University Bulletin, Series: Physics, V. 8, Iss. 3, P. 13 (2013) (in Russian).

[15] A.F. Buzulutskov, A.E. Bondar, A.D. Dolgov, A.V. Sokolov, L.I. Shekhtman, *TWO-PHASE CRYOGENIC AVALANCHE DETECTOR*, Russian patent RU 2517777 C2, Date of publication: 27.05.2014, Bull. № 15.

[16] A. Bondar et al., *Performance degradation of Geiger-mode APDs at cryogenic temperatures,* 2014 JINST 9 P08006 [arXive:1406.4633].

[17] G. Bondarenko et al., *Limited Geiger-mode microcell silicon photodiode: new results*, Nucl. Instrum. Meth. A 442 (2000) 187.

[18] E. Aprile et al., *Detection of liquid xenon scintillation light with a silicon photomultiplier*, Nucl. Instrum. Meth. A 556 (2006) 215.

[19] H. Otono et al., *Study of MPPC at Liquid Nitrogen Temperature*, PoS (PD07) 007.

[20] P.K. Lightfoot et al., *Characterisation of a silicon photomultiplier device for applications in liquid argon based neutrino physics and dark matter searches*, 2008 JINST 3 P10001.

[21] J. Haba, *Status and perspectives of Pixelated Photon Detector (PPD)*, Nucl. Instrum. Meth. A 595 (2008) 154.

[22] D.Y. Akimov et al., *Tests of multipixel Geiger photodiodes in liquid and gaseous xenon*, Instrum. Exp. Tech. 52 (2009) 345.

[23] A. Bondar et al., *Geiger Mode APD performance in a cryogenic two-phase Ar avalanche detector based on THGEMs*, Nucl. Instrum. Meth. A 628 (2011) 364 [arXiv:1003.1597].

[24] G. Collazuol et al., *Study of silicon photomultipliers at cryogenic temperatures*, Nucl. Instrum. Meth. A 628 (2011) 389.

[25] J.J. Csathy et al., *Development of an anti-Compton veto for HPGe detectors operated in liquid argon using silicon photo-multipliers*, Nucl. Instrum. Meth. A 654 (2011) 225.

[26] http://www.hamamatsu.com

[27] http://www.cpta-apd.ru, http://www.photonique.ch.

[28] http://www.sensl.com

[29] C.A.B. Oliveira et al, *Simulation of gaseous Ar and Xe electroluminescence in the Near Infra-Red range*, Nucl. Instrum. Meth. A 722 (2013) 1.

[30] P. Lindblom and O. Solin, *Atomic infrared noble gas scintillations I. Optical spectra*, Nucl. Instrum. Meth. A 268 (1988) 204.